\documentclass[11pt,a4paper]{article}

\usepackage{fontspec}
\usepackage[english]{babel}
\usepackage[margin=25mm]{geometry}
\usepackage{graphicx}
\usepackage{booktabs,tabularx,placeins}
\usepackage[protrusion=true,expansion=false]{microtype}
\usepackage[font=small,labelfont=bf,skip=6pt]{caption}
\usepackage{xurl}
\usepackage[hidelinks]{hyperref}

\AddToHook{cmd/section/before}{\FloatBarrier}
\title{Injectra: Retrofitting Extensibility into Web-Based Desktop Applications}
\author{Qin Sun\\\texttt{kingsun22515@gmail.com}}
\hypersetup{
  pdfauthor={Qin Sun},
  pdftitle={Injectra: Retrofitting Extensibility into Web-Based Desktop Applications},
  pdfsubject={Human-Computer Interaction}}
\date{}

\begin{document}
\maketitle

\begin{abstract}

Large language models can generate code based on natural language prompts, opening up new possibilities for users to customize software and extend its functionality. However, applying generated code to existing desktop software that users rely on daily still poses engineering challenges, such as integrating with heterogeneous host systems, adapting to runtime environments, and deploying enhanced features. Existing research has implemented various forms of software modification through specific interface toolkits, browser extensions, dedicated application architectures, and external overlay interfaces; however, their integration requirements and scope of applicability vary, and there remains a lack of a universal, cross-application extension solution for existing desktop clients.

This paper proposes and implements Injectra, a general-purpose code injection and feature enhancement system for existing desktop clients, suitable for applications whose user interfaces are built using embedded browser engines (such as Electron) or system WebViews (such as Tauri). Injectra decouples the host integration logic from the upper-layer enhancement code. Combining both runtime integration and pre-launch bootstrap mechanisms, it injects JavaScript, HTML, and CSS into the application’s internal Web runtime environment through a unified execution channel, thereby enabling modifications to interface rendering, adjustments to interactive behaviors, and client-side functionality extensions. The system uses a command-line interface (CLI) as a unified entry point, allowing external tools, AI agents, and MCP integration layers to reuse its injection and enhancement capabilities.

We tested 38 desktop applications already installed on our macOS machine; code injection was successfully completed for all of them. We demonstrated the system’s capability to extend functionality through various enhanced use cases and also successfully validated the Windows version. Based on Injectra, we further developed an LLM-powered enhancement tool driven by selection and annotation: users select interface objects within the original application and describe their modification requirements; the large language model then generates the enhancement code, which is injected, executed, and rendered by Injectra.

Experimental results demonstrate that Injectra possesses generalized, cross-application integration capabilities within typical desktop applications, and that various upper-layer enhancement tools can reuse its unified execution infrastructure. Injectra provides a unified in-application execution foundation for both manually written and AI-generated enhancement code, driving the evolution of existing desktop software from a closed model—which only provides predefined functions and customization options—toward a user-extensible software environment. The system's source code has now been released as open source: \url{https://github.com/sqmw/injectra}.

\end{abstract}

\section{Introduction}
\label{sec:introduction}

Users’ actual needs are often more specific than what a software’s preset features can cover, and they constantly evolve depending on the task and usage context. Within the same application, some users may want to hide controls unrelated to the current task, others may need to reorganize interface information, and still others may wish to add annotation, filtering, or auxiliary processing functions alongside existing content. At the same time, users have accumulated data, operational habits, and workflows through their daily use of software. Redeveloping an application or migrating to other tools to meet these needs entails significant costs; in contrast, directly supplementing the required capabilities within the user’s familiar software environment offers more immediate practical value. This is precisely the focus of software plasticity and end-user extension: enabling software to continuously adapt to specific usage contexts, rather than being limited to the configuration options provided in advance by developers \cite{maclean1990tailorable,ko2011euse}.

Large language models offer new technical opportunities for achieving this goal. Users can describe their modification intentions using natural language, with the model assisting in generating or adjusting implementation code, followed by iterative feedback based on the interface results. Stylette explores an interactive approach for expressing web styling goals through natural language, while DirectGPT combines direct manipulation with large language models, demonstrating the value of pointing to specific objects and making modification requests \cite{kim2022stylette,masson2024directgpt}. These studies indicate that a more direct connection can be established between users expressing their needs and modifying software. However, generating enhanced code is only one step in the process: the code must also be integrated into the application the user is currently using, executed within the correct interface context, and work in tandem with existing functionality.

This deployment challenge is particularly pronounced in existing desktop applications. Desktop clients typically organize their interfaces and functionality around their own product goals; whether and how external code can enter the application depends on the extension interfaces, debugging capabilities, and runtime environment provided by the host. Even if both clients use Web technologies to build their interfaces, their page discovery methods, launch processes, and code execution entry points may differ. If every enhancement tool must handle these differences anew, integration with the host would become coupled with specific functional implementations, and cross-application reuse would be limited. Therefore, a common integration and execution layer—one that can be reused by different enhancement tools—is needed to bridge the gap between natural language requirements and in-application functional extensions.

Research on software transformation has explored various implementation paths. For example, Jelly generates interfaces mediated by a task-driven data model and transforms users’ natural language commands and direct manipulations into modifications to this model, enabling the interface and data to continuously evolve according to task requirements \cite {cao2025malleable}; DOMSteer directly modifies the DOM in real-time web pages via browser extensions to provide users with in-situ assistance \cite{hao2026domsteer}; Interface Quilt, meanwhile, combines a computer operation agent with generative interfaces to overlay interface patches on existing applications, altering their presentation, interaction, and usability \cite{qiu2026interfacequilt}. These approaches demonstrate that existing software can gain new interactions and functionalities at different levels, but they correspond to different integration prerequisites: altering the data model within generative systems, modifying pages within the browser, and overlaying new interfaces externally—each acting on distinct software layers. This paper further focuses on the internal execution environment of existing Web UI desktop clients, investigating how to organize their integration capabilities into a reusable, general-purpose infrastructure.

Web UI desktop applications provide a common technical foundation for researching this issue. The mainstream architectures of such applications can be divided into two categories based on the source of the rendering engine: those that bundle a browser kernel within the application (such as Electron) and those that call the operating system’s WebView component (such as Tauri). Although they differ in how they integrate with the desktop host and platform, their Web interfaces share execution and rendering mechanisms based on JavaScript, HTML, and CSS. This commonality implies that the host-related integration process can be separated from the upper-layer enhancement logic: the former is responsible for establishing an execution channel for the target application and pages, while the latter utilizes the interface environment to implement the required functionality. Based on this observation, this paper proposes a research question: \emph {How can a universal code integration and enhancement execution layer be established for existing Web UI desktop applications, enabling both manually written and model-generated enhancement code to be deployed and executed within the original application, and reused by different upper-layer tools? }

Answering this question requires addressing challenges at both the integration and reuse levels. At the integration level, it is necessary to connect the execution entry points of different hosts to specific application pages, while accommodating both applications that are already running and those that require entry points to be prepared before startup. At the reuse level, upper-layer tools must be able to submit enhanced code, update enhanced content, and obtain execution results in a consistent manner, without having to implement integration processes for various hosts themselves. The universality here is reflected in the reuse of integration mechanisms and execution capabilities; object selection and business logic dependent on the semantics of specific pages are handled by the corresponding enhancement tools.

This paper proposes and implements Injectra, a general-purpose code injection and functionality enhancement system for existing Web UI desktop applications. Injectra decouples host integration logic from the enhancement execution mechanism and combines two paths—runtime integration and pre-launch initialization—to inject JavaScript, HTML, and CSS into the target application’s Web interface environment. On top of a shared execution mechanism, enhancement code can alter interface presentation, adjust interactive behavior, or extend client-side functionality. The system provides a unified command-line interface (CLI) and a reusable library layer, enabling enhancement payloads and external tools to share the underlying integration capabilities. This organizational approach allows Injectra to support enhancements from various sources, including scripts written by developers and code generated by large language models.

\begin{figure}[htbp]
\centering
\includegraphics[width=\linewidth]{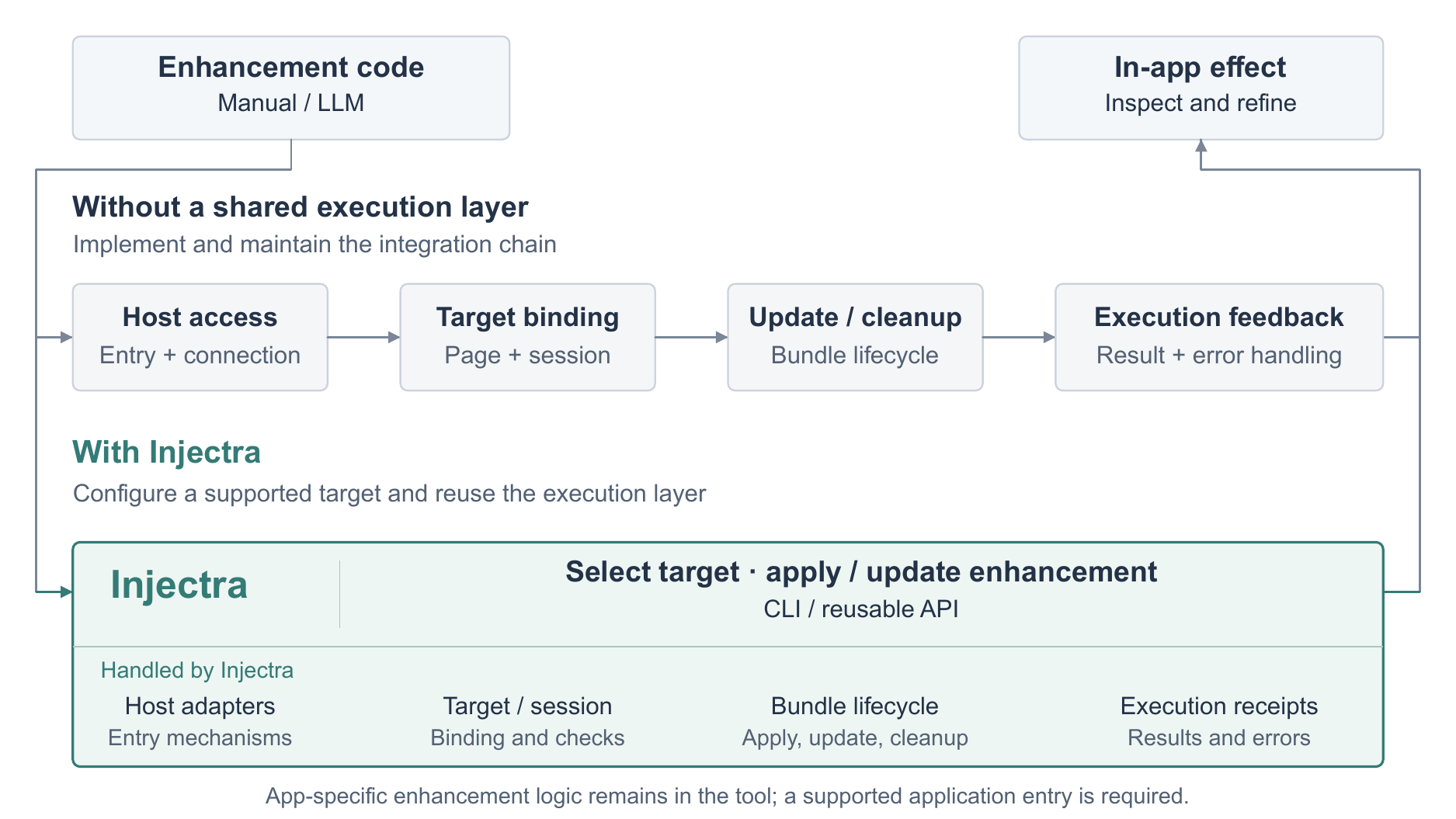}
\caption{Comparison of integration efforts when building personalized enhancement tools. In the top approach, the implementers of upper-layer tools organize and maintain the complete execution chain themselves; in the bottom approach, these capabilities are reused through Injectra’s unified entry point, reducing the integration steps that upper-layer tools need to implement. The mechanisms listed at the bottom of the green box are handled by Injectra and do not constitute additional operational steps. Both approaches share the same enhancement code and verify results within the original application, provided the target application has an available entry point. }
\label{fig:customization-workflow}
\end{figure}

Figure \ref{fig:customization-workflow} compares the workload required of upper-layer enhancement tools when building an execution layer from scratch versus using Injectra. The former requires managing host integration, target binding, payload lifecycle, and execution feedback; the latter, upon meeting the corresponding entry conditions, reuses these capabilities through unified calls, reducing the upper-layer integration process to target selection and payload submission, eliminating the need to reimplement the underlying execution chain. Both workflows can utilize manually written or model-generated code and verify effectiveness within the original application; the selection of page objects and business adaptation remain the responsibility of the enhancement tool.

Based on Injectra, we further developed an LLM enhancement tool driven by selection and annotation. Users directly select interface objects within the original application and describe their modification requirements; the tool then passes the corresponding context to the model to generate enhancement code, which is subsequently injected, executed, and presented via Injectra. This example places requirement capture and code generation atop a generic injection layer, demonstrating how upper-layer interactive tools can reuse underlying execution capabilities. This paper also examines the system’s cross-application feasibility and scope of functional expression through integration validation with multiple real-world desktop applications and various types of enhancement examples; specific test conditions, coverage, and results will be detailed in the evaluation section.

The main contributions of this paper include:
\begin{enumerate}
\item For existing Web UI desktop applications, we propose an architecture that separates host integration from upper-layer enhancement, enabling cross-application integration and enhancement execution to become independently reusable capabilities.
\item We have implemented the Injectra system, which combines runtime integration with pre-launch guidance. Through a shared execution mechanism and a unified invocation entry point, it supports the enhancement of in-application interfaces, interactions, and client-side functionality.
\item We provide feasibility validation and reference enhancement examples across multiple real-world applications, and demonstrate the complete integration between requirement expression, code generation, and in-application execution using an LLM tool driven by box-selection annotations.
\end{enumerate}

The following sections first review relevant research and define the system’s scope and design objectives, then introduce Injectra’s architecture and implementation, and subsequently demonstrate its capabilities through reference enhancement examples and system validation. Finally, we discuss its applicability limits and future research directions.

\section{Background and Related Work}
\label{_sec:background-related_}

\subsection{Background}
\label{_sec:background_}

\subsubsection{Web UI Desktop Applications}

A Web UI desktop application refers to a desktop client whose user interface is built, in whole or in part, using Web technologies. Such applications typically consist of two components: a native layer and a Web interface environment. The native layer handles window management, process organization, and calls to operating system capabilities, while the Web interface environment is responsible for executing and rendering documents, scripts, and styles. In terms of how the rendering engine is deployed, Electron integrates Chromium and Node.js into its application framework and distinguishes between the main process and the rendering process: the main process manages application windows, while the rendering process hosts the web pages within those windows \cite{electronprocess}. Tauri, on the other hand, uses the WebView provided by the operating system to render the front end and connects the front end to native capabilities through a communication mechanism between the Rust core and the WebView \cite{tauriarchitecture}. A single desktop application can contain multiple windows or WebViews, and a single window can also contain multiple documents and execution contexts.

The environments of the native layer and page scripts are not identical. For example, in Electron, the main process, preload scripts, and page scripts have different responsibilities, and context isolation can be applied between preload scripts and web pages \cite{electronprocess}. Whether a page can access files, window management, or other native capabilities depends on the interfaces exposed by the native layer to that environment. Therefore, when describing where code is executed within an application, it is necessary to distinguish between the application process, the WebView, and the specific page context.

\subsubsection{Web Documents and Execution Contexts}

HTML describes the document structure, CSS describes style rules, and JavaScript executes logic, handles events, and manipulates the document through interfaces provided by the runtime environment. The Document Object Model (DOM) represents a document as a node tree and defines nodes, events, and their operation interfaces \cite{whatwgdom}. Modifications to nodes or styles can alter the page’s rendering, while event-handling code links user actions to page behavior. Desktop windows, DOM nodes, and internal business objects within an application belong to different levels of representation: the nodes and properties accessible to a page are not directly equivalent to the host’s entire internal state.

The execution context specifies the global objects and variable environment visible to the script at runtime. A single page can have mutually isolated scripting environments: for example, content scripts in Chrome extensions are by default in an isolated environment, where their JavaScript variables are separated from the page’s scripts, yet they can still manipulate the corresponding document \cite{chromecontentscripts}. The debugging interface also organizes code execution through target and context identifiers; the Runtime domain of the Chrome DevTools Protocol (CDP) provides protocol operations related to expression evaluation and execution contexts \cite{cdpruntime}. Therefore, sending code to a JavaScript engine and executing a piece of code within a specific application page that can access the target document are two distinct operations that require separate descriptions.

\subsubsection{Code Injection and Enhancement Payloads}

This paper refers to the client that hosts the original interface and functionality as the \emph{ host application }, and refers to external code and resources that add rendering, interaction, or client-side logic to it as \emph{ enhancement payloads }.
\emph{ connecting to } refers to establishing a communication channel directed at the target host and its pages, while \emph{ injecting } refers to loading the payload into the target execution environment through that channel.
In a web interface, an enhanced payload can include JavaScript, HTML, and CSS; HTML and CSS are used as document and style resources, respectively, while associated scripts are responsible for assembling them into the page and implementing the corresponding behavior. The payload may originate from manually written files or code generated by other tools.

To describe the timing of establishing an access point, this paper uses the terms \emph{ “runtime access” } and \emph{ “pre-launch bootstrap” }. Runtime access targets applications that are already running and establishes a page execution channel through available debugging interfaces or host bridges. Pre-launch bootstrapping, on the other hand, first prepares a bootstrap code segment that is loaded when the application starts; this bootstrap code then connects to an external control endpoint and loads subsequent enhancement payloads. Both describe the timing and method of entry point establishment; the enhancement code subsequently runs within the execution environment of the corresponding application. Entry point establishment, payload generation, and payload updates are distinct stages in this process.

\subsubsection{End-User Extension and LLM-Assisted Modification}

End-user programming emphasizes programming activities that serve the user’s own tasks; it is defined by the purpose of the activity rather than the participant’s professional status or programming proficiency. End-user development also involves modifying and extending software during use \cite{ko2011euse}. Software plasticity focuses on the modifiability of software during use, such as when users continuously adjust its presentation, interaction, or functionality \cite{klokmose2015webstrates}. In the LLM-assisted modification process, natural language descriptions, the selection of interface objects, or direct manipulation can serve as inputs for modification intent, with the model generating or adjusting the corresponding implementation \cite{kim2022stylette,masson2024directgpt}. Requirement expression, implementation generation, and code execution correspond to different responsibilities in this process; it should be noted that LLM-based augmentation tools do not necessarily adopt an agent architecture: their code generation can be organized as a fixed request–generation–execution workflow without the need to introduce autonomous planning or multi-step decision-making mechanisms.

\subsection{Related Work}
\label{_sec:related-work_}

\subsubsection{Malleable Software Environments}

One line of research incorporates plasticity into the organization of documents, data, and applications by designing software environments that support sustainable modification. Webstrates treats web pages as shareable dynamic media; through DOM synchronization between the server and client, it enables collaborative modification of document content, interfaces, and code \cite{klokmose2015webstrates}. Its plasticity is built upon Webstrates’ hosted and shared environment; applications or documents gain the corresponding modification capabilities upon joining this environment. Jelly, on the other hand, generates interfaces mediated by a task-driven data model and translates natural-language requests and direct manipulations into modifications to the data model, causing the interface and data to update continuously\cite{cao2025malleable}. It organizes continuous customization as an evolutionary process of the data model and interface within the generation system.

These works provide environments and representation methods for building plastic software, but their modification mechanisms rely on corresponding shared foundations or generative models. When the research subject is a third-party client that has already been deployed but was not built using these environments, the question of how to integrate into its existing execution environment requires separate consideration. Injectra focuses on this integration layer: while preserving the existing application as the execution environment, it establishes reusable execution channels for external enhancement code. This objective complements the redesign of malleable software environments and does not require the uniform conversion of existing business data into the data model of any specific generative system.

\subsubsection{Browser-Based Automation and Web Augmentation}

Research on web automation and user scripts has demonstrated that clients can continue to modify the content and behavior of web pages after they have been rendered. Chickenfoot embeds a programming environment within Firefox and, by utilizing operations such as keyword matching targeting visible page components, supports web automation, customization, and cross-site integration \cite{bolin2005chickenfoot}. Therefore, adding functionality to existing pages via scripts is not limited to visual adjustments, nor does it require access to the website’s server-side source code. In recent years, DOMSteer has further integrated user requests, real-time interface context, and LLM processing to perform DOM insertion, modification, and reorganization via a Chrome extension, providing supplementary information and interactive support to the original webpage \cite{hao2026domsteer}. Its implementation reuses the page access and execution infrastructure provided by browser extensions.

The prerequisite for this approach is that the corresponding browser can load and run an extension or scripting environment; page matching, script loading, and permission management are also organized by this environment \cite{chromecontentscripts}. Although Web UI desktop clients also have a Web execution environment, their processes, windows, and code entry points are managed by their respective hosts; therefore, one cannot assume that they share the same extension loading mechanism merely because they share a DOM. Injectra therefore separates the integration of heterogeneous desktop hosts from upper-layer enhancements and investigates how to provide an execution foundation for browser page enhancement approaches tailored to existing desktop clients. This difference at the system level shifts the focus of this paper to the reuse of integration mechanisms across desktop hosts, whereas DOMSteer focuses on the auxiliary generation and delivery based on real-time web page contexts.

\subsubsection{Runtime Modification of Existing Desktop Applications}

There are also direct research precedents for runtime modification of existing desktop software. Scotty, through runtime toolkit overrides, enables developers to modify the interactions and behavior of existing Cocoa applications without accessing the original application’s source code \cite{eagan2011scotty}. The team’s report describes hooks for altering control rendering, event funnels for modifying event handling, glass sheets for overlaying interactions, as well as abstractions such as dynamic code loading and object delegation \cite{insitu2013scotty}.

It should be noted that this capability is not universally available in arbitrary desktop software but is closely tied to the runtime characteristics of Cocoa. The dynamic code loading, object delegation, and event interception relied upon by Scotty are built upon the dynamic object model and message dispatch mechanisms of the Cocoa and Objective-C runtimes; the host framework itself provides the ability to query and replace classes, methods, and event-handling behaviors at runtime. For native applications lacking such dynamic mechanisms, similar runtime modifications typically require binary instrumentation or reverse engineering techniques (such as dynamic instrumentation tools like Frida), which significantly increase compatibility, stability, and maintenance costs, thereby limiting the development and widespread adoption of unified abstraction frameworks. Therefore, what Scotty demonstrates is that, provided the host offers sufficient dynamic runtime support, interaction and functional extensions can be achieved without source code; this does not imply that runtime integration is equally feasible for any desktop software.

This environmental boundary constitutes a key condition for comparison in this paper: modification abstractions built upon the object models, event dispatching, and runtime mechanisms of specific native toolkits are difficult to directly migrate to hosts that lack these mechanisms. Injectra targets Web UI desktop applications. By leveraging the standardized execution foundation of shared page scripts, documents, and styles, it uses integration and adaptation to handle host differences, enabling upper-layer enhancements to reuse common Web execution capabilities. Building on this foundation, it combines runtime integration with pre-launch guidance to organize reusable enhancement capabilities. This paper further investigates precisely this cross-host Web execution layer and its reuse mechanisms.

\subsubsection{Natural Language and Direct Manipulation for Modification}

Another class of research focuses on the expression of modification intent and the transformation from user intent to specific modifications. Stylette converts users’ natural-language styling goals for web page components into suggested CSS properties and values, and supports iterative adjustments through a candidate panel \cite{kim2022stylette}. DirectGPT introduces interaction principles such as object selection, direct manipulation, reusable prompt operations, and undo into the LLM interface, supporting modifications to text, code, and vector graphics within its editing environment \cite{masson2024directgpt}. These systems contribute to the interpretation of styling intent and the organization of human-computer interaction, respectively, rather than focusing on the unified integration of heterogeneous desktop hosts.

These interaction mechanisms can provide a means of inputting requirements for enhanced tools, but the specific generation process does not automatically provide a code deployment channel for third-party desktop applications. The selection-and-annotation tool presented in this paper inherits the interactive approach of pointing to objects and expressing intent, serving as a reference application built on top of a generic injection layer. Accordingly, Injectra’s contribution lies in the foundation of integration and execution that can be shared by various higher-level tools, while the selection-and-annotation example demonstrates how this layered approach connects the expression of requirements with in-application augmentation.

\subsubsection{External Interface Patches and Pixel-Based Access}

Unlike entering an application’s internal execution environment, the external overlay approach alters how users interact with existing software by superimposing a new interaction layer on top of the target application. Research on this approach predates the emergence of generative tools. User Interface Fa{\c{c}}ades enables the adjustment, reorganization, and recombination of existing graphical interfaces through direct manipulation \cite{stuerzlinger2006facades}. Prefab, on the other hand, identifies interface components based on pixels, providing a foundation for implementing new interactive behaviors in existing interfaces \cite{dixon2010prefab}. These approaches, which start from interface fragments and visible pixels, respectively, demonstrate that modifying existing software is not limited to accessing internal application document objects. Interface Quilt uses a computer-operated agent to explore the interface, after which a generative tool constructs interface patches that overlay the original application; a runtime control layer is responsible for transmitting interface states and input operations between the patches and the target application \cite{qiu2026interfacequilt}. This approach does not rely on the target application’s source code, DOM, or pre-provided business APIs, enabling modifications to applications that employ different interface technologies.

At the same time, this external overlay approach introduces a series of additional costs. First, interface exploration and patch generation rely on multiple rounds of perception, reasoning, and manipulation by the agent, resulting in relatively high computational and interactive overhead. Second, the patch itself constitutes an independent overlay interface; its design requires reorganizing controls and interaction flows and is predicated on an understanding of the target application’s interface structure and task logic—this places an additional interface design burden on users or generative tools. Third, user operations do not act directly on the original application; instead, they are first received by the patch layer, then simulated and forwarded to the target application, while the original application’s state must be fed back to the patch layer to maintain synchronization. While this intermediate layer provides integration capabilities, it also increases the complexity of input consistency, interaction latency, and state synchronization; the original paper also lists patch construction, interaction verification, and state synchronization as major limitations \cite{qiu2026interfacequilt}.

Injectra utilizes the application’s internal Web execution environment as the enhancement vehicle, allowing direct access to the page’s DOM interface and JavaScript execution capabilities. The enhanced interface is embedded within the original page; users continue to interact within the original application, and input events are handled by the page’s own event mechanism. This eliminates the need for simulation and forwarding via an external patch layer, as well as the need to maintain state synchronization between two sets of interfaces. The trade-off of this design is the necessity of establishing a usable internal access channel, and its applicability is constrained by the target application’s Web execution environment. Consequently, the two approaches use visible interfaces and internal execution environments, respectively, as their primary access points, and carry out new interactions and functions externally and internally within the application: the former requires rebuilding the interface and forwarding input to support different interface technologies, while the latter relies on internal access to directly extend the interface and behavior within the original application.

\subsubsection{Positioning of This Work}

The aforementioned studies have established software extension capabilities at various levels, including malleable environments, web scripts, native toolkits, interaction modification, and external interface patching. Cross-application reuse, decoupling of code from the host, and AI-assisted modification constitute the common research threads among these works. Following this thread, this paper focuses its research question specifically on \emph {—specifically, the integration of generic code and enhanced execution } within existing Web UI desktop clients: by adapting to the host to handle different entry points, we connect runtime integration with pre-launch guidance to a shared page execution capability, which is then reused by various higher-level enhancements. The key to this approach lies in extracting the host integration work—previously scattered across specific enhancements—into an independent infrastructure, enabling page-level extensions to enter different clients through the same execution layer. Subsequent architectural explanations and real-world application validation will focus on the reuse of integration mechanisms, shared execution capabilities, and the boundaries of page semantic adaptation.

Table \ref{tab:related-work-positioning} summarizes the representative works mentioned above based on integration prerequisites, enhancement deployment locations, and research priorities. This comparison is intended to illustrate the software layers involved in different approaches and does not provide a unified ranking of efficiency, reliability, or coverage capabilities for which no paired measurements have been conducted.
\begin{table}[htbp]
  \centering
  \caption{ Prerequisites for integration, enhanced deployment locations, and research priorities of related work. The table provides a qualitative summary based on representative mechanisms and does not represent a paired performance evaluation or a comprehensive list of capabilities. }
  \label{tab:related-work-positioning}
  \small
  \setlength{\tabcolsep}{4pt}
  \renewcommand{\arraystretch}{1.2}
  \begin{tabularx}{\linewidth}{@{}>{\raggedright\arraybackslash}p{0.20\linewidth}>{\raggedright\arraybackslash}p{0.23\linewidth}>{\raggedright\arraybackslash}p{0.23\linewidth}>{\raggedright\arraybackslash}X@{}}
    \toprule
    Representative Work & Prerequisites for Integration & Enhanced Host Locations & Research Focus \\
    \midrule
    Webstrates / Jelly~\cite{klokmose2015webstrates,cao2025malleable}
      & Documents participate in a shared environment, or interfaces are generated by corresponding models
      & Shared DOM foundation or data models and interfaces generated by the system
      & Support for software plasticity and continuous evolution within the environment \\
    \addlinespace
    Chickenfoot / DOMSteer~\cite{bolin2005chickenfoot,hao2026domsteer}
      & Browsers capable of loading corresponding scripts or extensions
      & Web pages and their DOM within browsers
      & Web page automation and in-situ enhancement leveraging browser infrastructure \\
    \addlinespace
    Scotty~\cite{eagan2011scotty,insitu2013scotty}
      & Cocoa hosts with corresponding runtime mechanisms
      & Native windows, controls, objects, and event handling
      & Modifying existing desktop applications using runtime toolkits \\
    \addlinespace
    Stylette / DirectGPT~\cite{kim2022stylette,masson2024directgpt}
      & Editing interfaces provided by web extensions or research systems
      & Web styles or manipulable objects within the system
      & Expressing and completing modifications using natural language and direct manipulation \\
    \addlinespace
    UI Fa{\c{c}}ades / Prefab~\cite{stuerzlinger2006facades,dixon2010prefab}
      & Interface fragments can be rearranged, or interface components can be identified on a pixel-by-pixel basis
      & Enhancements to existing graphical interface fragments or those identified at the pixel level
      & Rearranging interactions or implementing new behaviors based on existing interfaces \\
    \addlinespace
    Interface Quilt~\cite{qiu2026interfacequilt}
      & Visible interfaces can be detected and manipulated by computer-based tools
      & Overlay interface patches outside of existing applications
      & Modifying existing applications by generating new interface structures \\
    \addlinespace
    Injectra (this paper)
      & Web UI desktop host capable of establishing internal page channels
      & Web execution environment within the original application
      & Separating host integration and enhancement logic to reuse a common execution layer \\
    \bottomrule
  \end{tabularx}
\end{table}

\section{Scope and Design Goals}
\label{_sec:scope_}

The central research question of this paper is: How can reusable code integration and enhanced execution capabilities be provided for existing Web UI desktop applications? This question involves both the execution entry points in different hosts and how upper-layer enhancements can reuse these capabilities once integration is established. This chapter first defines the research subjects, scope of capabilities, and system responsibilities, and then proposes design objectives to guide architectural division and technology selection.

\subsection{Research Scope}
\label{_sec:research-scope_}

\subsubsection{Target Applications and Execution Layer}

The research subjects are existing desktop clients that have been deployed and whose user interfaces are built using an embedded browser engine or the system’s WebView. Enhancements are applied to the Web pages within the application and their execution environments, extending their presentation and behavior through JavaScript, HTML, and CSS. For hybrid applications that contain both native controls and WebViews, this paper discusses only the Web interface portion; native windows or controls are not considered targets for enhancement simply because they belong to the same process as the WebView. Server-side implementation modifications, general native interface refactoring, and arbitrary process memory modifications are outside the scope of this paper.

Access channels can be established through various mechanisms, such as debugging protocols or host bridging, and may involve calls to the host’s native layer when necessary. The underlying mechanisms used to establish the channel and the level at which the enhancement code ultimately operates must be described separately. This paper organizes enhancement capabilities around a shared Web execution environment and does not require all hosts to adopt the same underlying access mechanism. In terms of platforms, the implementation and verification in this paper focus on macOS (Apple Silicon) and Windows; support for Linux is reserved for future work. Specific host versions and runtime conditions are detailed in the Implementation and Evaluation sections.

\subsubsection{Enhancement Capabilities and Access Paths}

The scope of functionality covers three levels: interface presentation, interactive behavior, and client-side logic. Presentation enhancements include adjustments to styling, layout, and information organization; interaction enhancements include adding operation entry points, modifying local interaction methods, and integrating new event-handling logic; client-side logic enhancements can involve adding annotation, filtering, or auxiliary processing functions to the content accessible on the page. These capabilities enter the original application via external enhancement payloads, which can be written by developers or generated by LLM-driven upper-layer tools. The DOM, page state, and interfaces accessible to the payload still depend on the execution environment actually provided by the target page and the host application.

This paper also incorporates two approaches: runtime integration and pre-launch initialization. Runtime integration targets an already running application, establishing a page execution channel through available entry points; pre-launch initialization first prepares an initialization entry point in a derived copy of the application, then loads and updates the external enhancements. Once the initialization entry point is prepared, page enhancements can be iterated independently of the application package, without requiring the application to be repackaged after each modification to the payload. The common goal of both approaches is to provide page execution capabilities to upper layers; the entry conditions and preparation processes are handled by the corresponding host application adapters.

\subsubsection{System and Enhancement Responsibilities}

Injectra is responsible for establishing page execution channels for the target application, accounting for differences in process discovery and application launch methods across operating system platforms such as macOS and Windows, as well as variations in debugging interfaces, WebView organization, and launch entry points among different host applications. Once the channel is established, Injectra takes full responsibility for identifying and selecting the target page, injecting and managing the execution of the enhancement payload, and providing feedback on execution results; the payload code itself continues to run within the target page’s Web execution environment.

The enhancement author or upper-layer tool is responsible for the specific functional content, including selecting which UI elements to target, interpreting page semantics, meeting user needs, and handling the business logic of the enhancement itself. Therefore, generality is reflected in the fact that the integration mechanism and execution capabilities can be reused across multiple applications and enhancements, without requiring that the same segment of code—which depends on specific page semantics—remain unchanged across all applications.

Annotation-driven LLM tools sit above this execution layer, responsible for capturing modification intent, organizing context, and generating enhancement code. Their role is to demonstrate the types of upper-level workflows that general injection capabilities can support; Model selection, prompt design, and agent orchestration are not part of the core responsibilities of injection. Similarly, the completion of page execution and the generation of code that satisfies the user’s intent are two distinct criteria: the former is supported by execution feedback and page effect verification, while the latter requires evaluation specific to the enhancement task.

\subsubsection{Operational Boundaries and Non-Goals}

Injectra executes enhancement code only within target applications explicitly specified and actively initiated by the user; the scope of this research does not include providing a secure sandbox for arbitrarily generated code. Scripts within a page possess the permissions allowed by their respective execution environments; selecting a specific area does not imply that the scope of the injected script’s impact will automatically be restricted to that area. Operations such as the removal of managed nodes and the cleanup of registered payloads only handle resources registered by Injectra; this does not imply that persistent data written by the script or its impact on the external environment will also be revoked.

This document focuses on targeted integration, execution, and enhancement updates. Automatic re-injection following page navigation, automatic cross-window enhancement recovery, or automatic adaptation of enhancements after application updates are not considered part of the current capability requirements. Page object targeting is implemented by specific enhancement payloads, while long-term adaptation following changes to the application or page is the responsibility of the enhancement payload author or higher-level tools. A complete workspace, built-in editor, plugin marketplace, and generic agent platform are also not included as component objectives of this system; existing editors, external tools, and model services may participate in the enhancement process by calling APIs. This boundary allows the research to focus on generic integration and execution reuse in real-world applications, while reserving research on long-term maintenance, code generation, and user interaction as composable higher-level problems.

\subsection{Design Goals}
\label{_sec:design-goals_}

Within the scope described above, the system design adheres to the following five objectives. These objectives provide a common basis for subsequent module division, integration paths, and interaction workflows, and are validated through corresponding system behaviors.

\subsubsection{G1: Cross-Host Generality}
\label{_goal:generality_}

The system should leverage the common execution foundation of Web UI applications and converge differences among operating system platforms, host applications, and integration mechanisms into adaptation implementations with clearly defined boundaries. This objective is adopted because if every new application required the complete reimplementation of the injection process, it would be difficult to establish a reusable infrastructure for integration. Therefore, new applications with identical technical conditions should prioritize reusing existing integration implementations; only when there are genuinely different host mechanisms or version requirements should corresponding adaptation be performed. Generic applicability is supported by clear, common interfaces and their reuse across multiple real-world hosts, rather than by writing application-specific branches within the core process.

\subsubsection{G2: Separation of Access and Enhancement Logic}
\label{_goal:separation_}

Host integration and upper-layer enhancements should have distinct responsibilities, so that enhancement authors do not need to reimplement low-level connections and execution entry points for each feature. Accordingly, the adaptation layer is responsible for handling differences between operating system platforms and specific host applications, and for establishing an execution channel to the target page; the shared execution layer is responsible for injecting enhancement payloads through this channel and managing their execution; while the page element targeting and business logic required for specific enhancements remain within the enhancement payload. This requires different integration routes to reuse a common enhancement execution mechanism, thereby avoiding the need to repeatedly implement the same payload processing logic for different operating system platforms and preventing the intermingling of specific application interface rules into the general integration process. This separation provides a common execution foundation for manual scripts, reusable enhancement packages, and model-generated code.

\subsubsection{G3: In-Application Iteration with Explicit Resource Ownership}
\label{_goal:iteration_}

Enhancements should be rendered and validated within the user’s original application, ensuring that modification feedback corresponds to the actual page and usage context. The system must therefore support the deployment, updating, and cleanup of managed resources associated with enhancement payloads, and ensure that these managed resources have identifiable ownership so that subsequent operations can locate their targets. For pre-launch onboarding routes, one-time setup should be separated from subsequent payload iterations, allowing users to modify external enhancements via existing editors and view the results in the live application. This objective emphasizes the iteration and resource management of enhancements within the original working environment; the resource attribution mechanism is used solely to identify and clean up enhancement resources registered with Injectra and does not imply that script behavior is isolated or that all of its effects can be rolled back.

\subsubsection{G4: Composable Access for External Tools}
\label{_goal:composition_}

General-purpose injection capabilities should be usable by multiple callers without being tied to specific editors, models, agent frameworks, or integration protocols. Therefore, the system must provide a stable entry point for programmatic calls, organizing target selection, payload submission, and execution feedback into operations that external tools can compose. The CLI and reusable library layers serve this purpose; access layers such as natural language interaction, box-selection annotations, and MCP can organize their respective workflows on top of them. The responsibility of the external interface is to expose callable injection and enhancement capabilities, while the specific methods by which callers generate code and schedule tasks remain independent.

\subsubsection{G5: Explicit Targets and Execution Outcomes}
\label{_goal:outcomes_}

Augmentation operations should target specific application instances and pages, avoiding the assumption that the correct execution environment has been obtained based solely on the application name or the existence of a process. Access capabilities, page identity, and session state must be verifiable conditions prior to execution; when targets change or capabilities are insufficient, recognizable results should be provided to the caller. Execution feedback must also distinguish between three scenarios: not executed, executed, and undetermined results. Notably, “executed” does not equate to “successful execution”; it may still be accompanied by script exceptions or produce only partial effects. For operations with uncertain outcomes, automatic retries should not be used to mask this uncertainty, as repeated execution may introduce new side effects. This objective provides upper-layer tools with interpretable execution boundaries, enabling them to determine subsequent interactions and verifications based on actual feedback.

The above objectives constrain different levels of subsequent design: G1 and G2 guide the division of the host adaptation and shared execution layers; G3 guides enhancement updates and resource management; G4 guides external invocation methods; and G5 guides target binding and result feedback. Corresponding verification requires observing whether integration implementations are reused, whether different enhancements can use a common channel, whether updates produce effects on the correct pages, and whether exceptions and target changes are clearly expressed. These observable conditions will link the subsequent architectural descriptions with the evaluation content; specific mechanisms and verification results are elaborated in the corresponding sections.

Table \ref{tab:design-goal-mapping} maps the five objectives to the architectural mechanisms and observable verification conditions in Section \ref{_sec:architecture_}, serving as the organizational basis for the subsequent evaluation.
\begin{table}[htbp]
  \centering
  \caption{ Correspondence between design objectives, architectural mechanisms, and test criteria. The test criteria are used to organize subsequent evaluations and do not indicate that the corresponding experimental results have been reported in this table. }
  \label{tab:design-goal-mapping}
  \small
  \setlength{\tabcolsep}{4pt}
  \renewcommand{\arraystretch}{1.2}
  \begin{tabularx}{\linewidth}{@{}>{\raggedright\arraybackslash}p{0.16\linewidth}>{\raggedright\arraybackslash}p{0.37\linewidth}>{\raggedright\arraybackslash}X@{}}
    \toprule
    Design Objectives & Corresponding Mechanisms (Section \ref{_sec:architecture_}) & Observable Test Criteria \\
    \midrule
    \textbf{G1}\newline Cross-Host Portability
      & Isolate platform, host, and entry point differences within the access adaptation
      & Applications with identical technical conditions can reuse the access implementation without adding brand branches to the core \\
    \addlinespace
    \textbf{G2}\newline Decoupling Access from Enhancement
      & Common Web Contract, Bundle Assembly, and Single-Source Page Runtime
      & Different backends use the same payload generation mechanism; page selectors and business logic remain within the enhancement \\
    \addlinespace
    \textbf{G3}\newline In-Application Iteration and \newline Resource Ownership
      & Register nodes and cleanup callbacks by enhancement identifier; separate entry point preparation from payload updates
      & Updates within the same page replace managed resources and render new effects; cleanup scope is verifiable \\
    \addlinespace
    \textbf{G4}\newline Combination of external tools
      & CLI and reusable library layers provide entry points for targets, submissions, and feedback
      & Manually written bundles and external enhancement tools can reuse the same set of invocation capabilities \\
    \addlinespace
    \textbf{G5}\newline Clear objectives and results
      & Backend-specific target references, session/capability verification, and deterministic execution feedback
      & Operations are rejected if targets are incorrect or capabilities are insufficient; operations are not automatically replayed when results are unknown \\
    \bottomrule
  \end{tabularx}
\end{table}

\section{Injectra Architecture}
\label{_sec:architecture_}

Section \ref{_sec:scope_} proposes design goals for cross-host access, enhanced execution, and the separation and reuse of upper-layer functions. Based on these goals, Injectra encapsulates host-related entry mechanisms as access adapters, organizes the invocation methods of target pages into a common contract, and executes external enhancement payloads through a shared runtime. This section explains this architecture in terms of component responsibilities, target representation, access paths, payload lifecycle, and tool composition; specific platform mechanisms and implementation requirements will be discussed in the next section.

\subsection{Architecture Overview}
\label{_sec:architecture-overview_}

\begin{figure}[htbp]
\centering
\includegraphics[width=\linewidth]{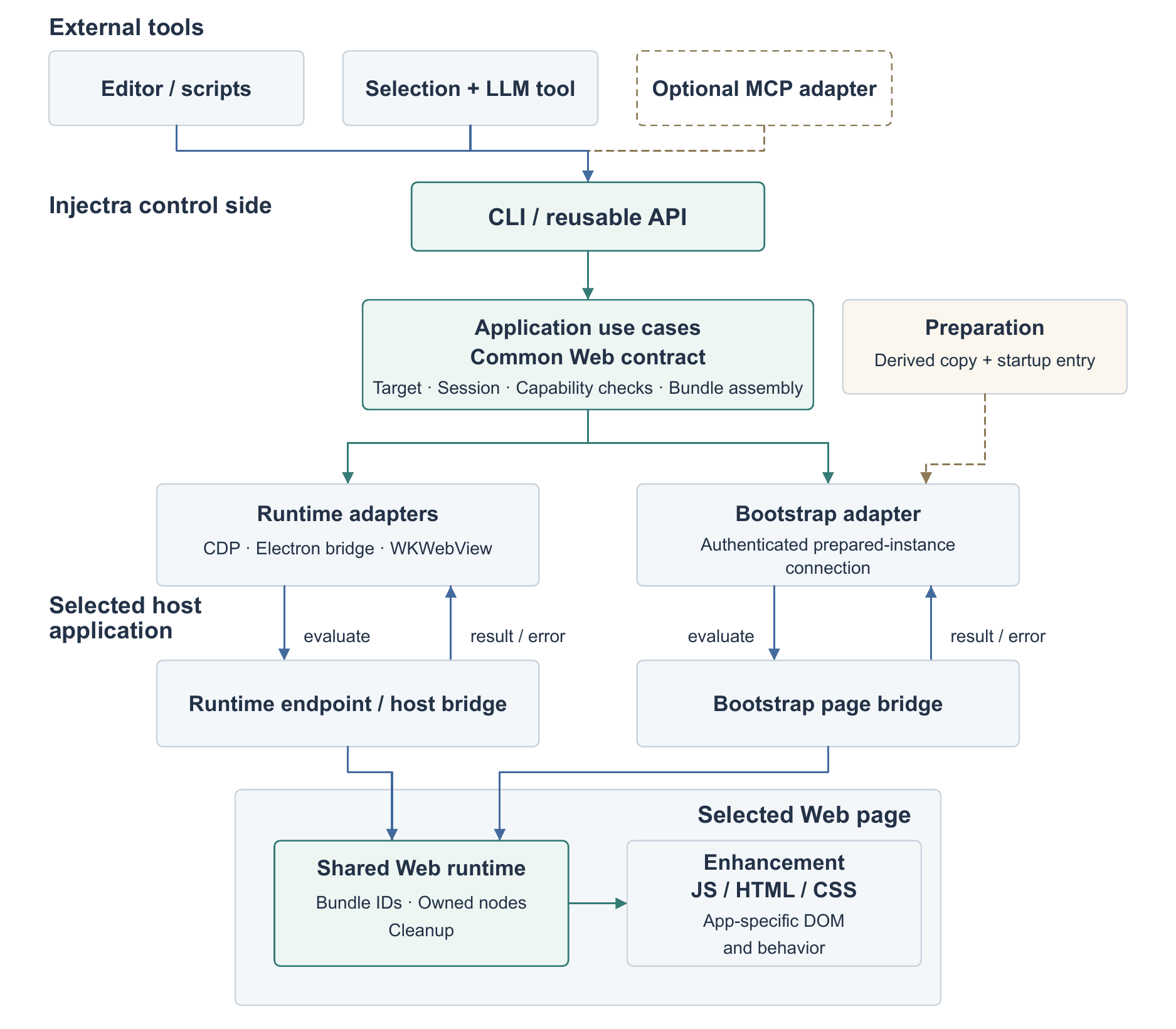}
\caption{Injectra’s logical architecture: Two integration paths connect to the selected page via a common Web Execution Contract and reuse a shared runtime within the page. Solid lines represent calls and feedback; dashed lines represent startup preparation or optional external integration; MCP is used solely to illustrate the location of external composition.}
\label{fig:architecture-overview}
\end{figure}

Figure \ref{fig:architecture-overview} illustrates the system’s logical organization. The components in the figure represent their respective responsibilities, and the lines indicate the primary interaction relationships between components rather than dependencies between source code modules; solid lines denote runtime calls and result feedback, while dashed lines denote pre-launch initialization or optional external integration. The overall structure distinguishes between the control end and the target application: the control end accepts calls, assembles payloads, and organizes integration, while the web pages within the target application host the shared runtime and specific enhancement code. The control end submits programs and receives return values or errors, whereas the page end generates actual effects through its own DOM, styles, and event mechanisms. Therefore, the fact that the control end can initiate execution does not imply that the payload logic runs on the control end.

The external invocation entry point consists of the CLI and a reusable library layer. The CLI handles parameter parsing, specific backend assembly, and result output; the library layer provides use cases such as target discovery, application enhancement, enhancement removal, and page execution, enabling external programs to invoke these same capabilities within their own workflows. These use cases rely on an abstract Web execution contract rather than directly depending on any specific debugging protocol or host bridging implementation. Specific adaptation can be handled by the CLI or provided by external programs. Regardless of the approach, use cases follow the same processing flow: first, verify target ownership and page execution capabilities; then, submit the program to be executed to the selected page—without needing to alter their own implementation based on changes in the underlying access mechanism.

The integration adapter is responsible for mapping the execution entry points actually provided by the host to this contract. Platform differences, host mechanisms, and page functionality are thus handled at different levels: the platform and host adapters handle the application environment and execution entry points; the shared layer handles payload execution; and the enhanced code handles page semantics and specific business logic. Injectra’s core layer uniformly defines only three types of information: the identification method for target pages, the description method for session capabilities, and the meanings of different error types. Whether the target page provides a DOM, how scripts are executed within the page, and what transmission protocol is used between the control end and the page are not within the scope of the core layer’s specifications; instead, they are handled separately by the Web Execution Contract and the specific integration adapter. This division clarifies the boundary between the core layer and Web execution capabilities, preventing any host-specific internal mechanism from being treated as a prerequisite that all integration implementations must satisfy.

\subsection{Targets, Sessions, and Capabilities}
\label{_sec:architecture-targets_}

Cross-host enhancement first requires ensuring that code is delivered to the page actually specified by the caller, rather than simply sent to an application or process with the same name. The existence of an application name or process can only be used to discover candidate targets; it cannot uniquely identify a page within a window or its execution environment. Injectra therefore uniformly represents connectable pages as Web targets, separating page identity from the implementation details of specific hosts, enabling upper-layer use cases to specify operation targets in a consistent manner.

Each Web target contains a target reference to the associated backend, as well as metadata such as the title, URL, and page context. A target reference consists of the backend instance’s identity and the target key interpreted by that instance; even if two backends return the same local target key, they cannot be regarded as the same target. This reference is valid only within the current controlling process and the backend instance; it is not a universal object handle that can be persistently stored across processes. This design preserves the host information needed to locate a page while preventing the misuse of local identifiers across different backends.

Building upon the unified target representation, the Web Execution Contract further separates target discovery, session establishment, and page evaluation. Backends list targets, adapters verify the corresponding entry conditions, and sessions are established for selected targets; sessions preserve bound targets, declare the operations they support, and provide an interface for page evaluation. Use cases verify the backend ownership of a target before establishing a connection, re-verify the target bound to the session after connection, and check whether the session possesses the page execution capabilities required for the current operation. Consequently, target selection, host access, and capability checks form a verifiable execution path, preventing target keys from other connections from being passed to the current backend or code from being submitted when the session lacks the necessary capabilities.

The common contract also unifies the semantics of page evaluation: code must be evaluated in an execution environment capable of accessing the target page’s DOM, wait for the returned Promise, pass the result back to the control end as a JSON value, and treat script exceptions as execution errors. The fact that a host can run JavaScript is not sufficient proof that it satisfies this contract; for example, code running in an isolated script engine may not be able to access the target page’s DOM. For Injectra, these constraints are key to enhancing cross-host reusability: higher-level tools deal with consistent page execution semantics, rather than having to handle execution locations, asynchronous results, and exception representations separately for each host. The current common goal is based on the main page context and does not automatically treat arbitrary frames, workers, or other isolated environments as execution objects with the same semantics.

A capability declaration describes the operational capabilities that an adapter confirms when establishing a session, allowing use cases to verify them before execution. It does not guarantee that the target page will always be available after the session is established; page destruction or connection failure may still cause subsequent execution to fail, which must be expressed through execution feedback. Capability declarations also do not restrict a script’s actual permissions within the page and therefore do not constitute a security sandbox.

Through the three layers of target, session, and capability declaration, Injectra translates “which page to operate on, through which connection to execute, and what can currently be executed” into explicit checks within a generic workflow. This allows different higher-level tools to reuse the same set of target binding and pre-execution verification mechanisms without needing to understand the page identifiers and connection details of each host, thereby laying the foundation for cross-host reuse in G1, the separation of integration and enhancement in G2, and the clear definition of targets and execution results in G5.

\subsection{Access Paths and Host Adaptation}
\label{_sec:architecture-access_}

Injectra separates the establishment of the entry point from the subsequent execution of enhancements, supporting two paths: runtime integration and pre-launch initialization. The two paths have different prerequisites, but once page execution capability is obtained, both utilize Web contracts to support the same enhancement use cases and share the same runtime. While the two paths share the semantics of target operations and enhancement execution, this does not imply that all hosts employ the same underlying technology.

Runtime integration targets applications that are already running and have an available entry point. Adapters are responsible for discovering specific pages, establishing connections, and routing evaluation requests to the corresponding execution environment. Existing implementations include CDP-based page integration and host bridging for Electron and WKWebView. These approaches vary in their reliance on debugging endpoints, host interfaces, or native runtime conditions; some bridges require passing through the host’s native layer to obtain execution capabilities for Web pages. The enhancements ultimately act upon the Web page, which is distinct from the underlying methods used to establish the channel.

Pre-launch bootstrapping is intended for hosts that require entry points to be prepared in advance. The preparer identifies the entry point layout, injects a small amount of bootstrap code into a derived copy of the application, and retains the preparation information needed for subsequent connections. After the application launches, the bootstrap code establishes an authenticated communication channel with the local control endpoint and, upon discovering a page, implements a shared Web contract through the bootstrap backend. One-time entry point preparation is thus separated from subsequent payload iterations: after modifying external enhancement files, new content can be submitted to the established page channel without having to rebuild the application package every time an enhancement is modified. This process reuses the preparation mechanism based on the local Web entry point, rather than application-specific custom logic; as long as the entry point resources can be located and modified, and the derived copy can launch normally, subsequent enhancements can reuse the same execution process. When the entry point is protected by integrity checks, or when the interface is loaded entirely from a remote source, separate adaptation or alternative access paths are required.

Adaptation is categorized based on reusable protocols, host mechanisms, and resource formats, rather than application names. Applications with identical technical conditions should prioritize reusing existing implementations; corresponding strategies are introduced only when differences in the entry point or host require independent maintenance. Specific adaptation is assembled by the system at the call boundary; use case and payload processing logic are not reimplemented across application brands. Therefore, cross-application extensions primarily involve adding target configurations, applicable conditions, and compatibility evidence; when new code is absolutely necessary, its dependencies are also restricted to the corresponding adaptation.

\begin{figure}[htbp]
  \centering
  \includegraphics[width=\linewidth]{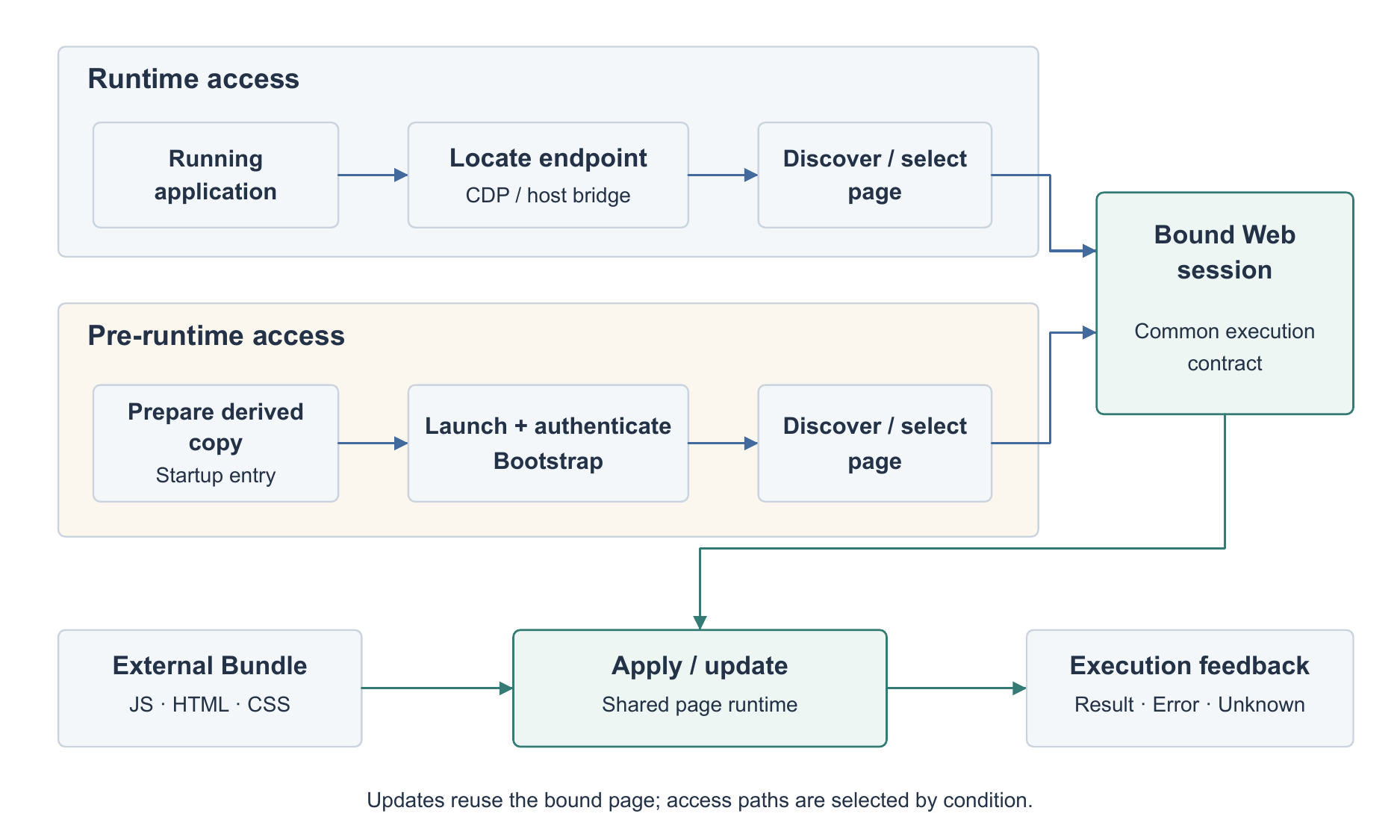}
\caption{ The two access paths and their subsequent shared execution flow. The pre-launch path first prepares and launches a derived replica; after page discovery and session binding, both paths reuse the payload application and result feedback. Arrows indicate logical steps and do not imply automatic selection, failure fallback, or repeated execution. }
  \label{fig:access-paths}
\end{figure}

Figure \ref{fig:access-paths} outlines two access paths step by step: runtime access uses existing entry points, while pre-launch guidance first prepares and launches a derived instance; after completing page discovery and session binding, both paths enter a common payload application and result feedback process. This temporal perspective complements the component division of labor shown in Figure \ref{fig:architecture-overview}; there is no automatic fallback relationship between the two paths.

\subsection{Enhancement Bundles and Shared Execution}
\label{_sec:architecture-bundles_}

Enhanced payloads are organized into Bundles, whose versioned manifests declare enhancement identifiers and optional JavaScript, HTML, and CSS resources; resource paths are resolved relative to the manifest directory. After the control end reads and validates these resources, it assembles them into a program ready for execution. Bundles describe only the deployment content; page selectors and business logic are implemented by the payload, while the host access method is determined by lower-layer adaptation.

All backends reuse the same Web runtime to assemble the program, avoiding the need to implement DOM creation and update logic separately. The runtime adds CSS to the page, inserts HTML into the created root container, and executes the payload script, while providing the script with a reference to the root container and a registration entry point for cleanup callbacks. If no HTML is provided, the root container may be empty, and the payload may directly manipulate the original page.

The shared runtime stores created nodes and cleanup callbacks in the page based on enhancement identifiers. When the same identifier is applied repeatedly, the runtime first cleans up old instances based on these records before assembling new content; when removing, it locates resources based on the same records and prevents overlapping execution on the same identifier. These records exist only within the runtime of the current page and do not handle cross-application or persistent plugin management.

Identifier-based cleanup is used to replace or remove individual enhancements without refreshing the page; it only overwrites registered nodes and callbacks. If the new payload fails to execute, the runtime attempts to clean up the resources created during this round and reports an error; however, previously cleared versions are not automatically restored. Refreshing the page further clears the runtime state in the current document, but it interrupts the original page task and cannot undo persistent data or side effects in the external environment; therefore, it is not equivalent to a full rollback.

\subsection{Iteration and Execution Feedback}
\label{_sec:architecture-feedback_}

During the dev process that boots the route before launch, the control end listens for changes to the Bundle manifest file and its declared resource files, and organizes the read content into versions and computes a digest identifier. If a file changes again during the reading process, that candidate version is discarded; stable versions are executed sequentially, and identical digests are not submitted repeatedly due to duplicate listening events. The target page is bound to the session; when the page is destroyed, rebuilt, or the connection fails, the current update cycle stops and does not automatically transfer to another page.

Execution feedback not only distinguishes between error types but also indicates at which stage the failure occurred and whether the code has already been executed. Specifically, \texttt{NotExecuted} indicates that the operation was rejected before submission, or there is clear evidence that it was not executed; \texttt{Executed} indicates that execution was confirmed to have been initiated on the target, but success of all statements is not guaranteed; \texttt{Unknown} indicates that the request may have been executed, but the result cannot be confirmed at this time. Script exceptions and return code failures are classified as post-execution errors; timeouts or disconnections cannot be used to determine that execution did not occur. A normal acknowledgment indicates that the program has completed evaluation; it does not imply that the enhanced functionality meets user requirements, and the caller must still perform subsequent acceptance testing based on the execution results.

To prevent repeated execution when the result is unknown, the dev process records the request identifier, content version, and target before submission, and saves the acknowledgment upon receiving a definitive result. When the result is \texttt{Unknown}, the system retains the pending confirmation status and stops automatic updates; after an abnormal termination of the control process, these unconfirmed records also prevent the system from automatically resubmitting the same request upon restart. The caller must verify the target status and explicitly confirm it before initiating a new execution. A single runtime call returns a result or an error directly, and the caller decides on subsequent handling.

\subsection{Composition with External Tools}
\label{_sec:architecture-composition_}

Injectra separates the generation of enhanced code from its execution within the page, enabling different external tools to reuse the same set of integration and execution capabilities. Developers can use existing editors to write Bundles; external programs can use the CLI or library layer to select targets, generate code, and update the enhancement, without needing to reimplement the host connection. Page semantics and specific business logic are still handled by the enhancement payload. Integration layers such as MCP can also be layered on top of the CLI or library layer to provide the same capabilities to other tools without modifying the page runtime.

Annotation-driven LLM tools that operate on selected text are examples of this approach. The requirement capture component records the selected area and annotations on the original page; the control end then organizes the context based on this information, invokes the model to generate or modify the enhancement code, and applies the candidate payload to the same page via Injectra. Since the capture component and the modified enhancements use different identifiers, annotation entry points are not removed when updates are applied, allowing users to continue proposing the next round of modifications. The tool is responsible for model selection, context organization, and ensuring that the generated results align with the user’s intent, while Injectra provides page integration, execution, and result feedback. The entire process can run according to fixed steps without requiring autonomous planning or agent orchestration within the injection core.

At this point, host adaptation, shared execution, resource management, a unified invocation entry point, and execution feedback have respectively implemented the aforementioned G1–G5. Subsequent implementation details and system evaluation will revolve around these mechanisms.
\FloatBarrier

\section{Implementation}
\label{_sec:implementation_}

This section describes the specific implementation of the architecture described in Section \ref{_sec:architecture_} within the current system. The implementation follows the component division shown in Figure \ref{fig:architecture-overview} and the two access paths shown in Figure \ref{fig:access-paths}, focusing on how the host entry point is transformed into page execution capabilities, how enhanced payloads are assembled, and how page changes and exceptions are identified. Application coverage, enhancement instances, and evaluation results are reported in the next section.

\subsection{Software Organization and Interfaces}
\label{_sec:implementation-organization_}

The control plane of Injectra is implemented in Rust and is divided into four layers: common contracts, Web execution, application use cases, and specific backends. Upper-layer use cases depend only on the \texttt{WebBackend} and \texttt{WebSession} interfaces and do not directly reference specific protocol implementations; therefore, they can run on different access backends. Target binding and capability checks are unified at the library layer, allowing both the CLI and external programs that directly call the library to reuse the same execution workflow.

The CLI provides commands for target discovery, application enhancement, removal of enhancements, and \texttt{prepare}/\texttt{dev}, among others. It selects and initializes the appropriate backend based on command parameters, then passes it to the library-layer use case for execution. Successful results of ordinary one-time operations are output as JSON; continuous development workflows output JSON events line by line; and fatal errors are written to the standard error stream, making it easy for external programs to distinguish between state changes, execution receipts, and command failures. Backends are selected based on call parameters; in the event of failure, the integration path is not automatically switched. Model invocation, interface object selection, and specific functional payloads are maintained by external tools or enhancement authors and are not the responsibility of the CLI integration use case.

The shared page runtime is implemented in JavaScript; it executes enhancement code within the target page and manages related resources. Host bridging and pre-launch preparation are implemented via auxiliary Python and JavaScript programs, which are uniformly invoked by the Rust control end.

\subsection{Runtime Access Backends}
\label{_sec:implementation-runtime_}

\subsubsection{CDP Page Access}

The CDP backend retrieves a list of debuggable pages from a specified loopback port and treats pages that provide a WebSocket endpoint for page debugging as candidate targets. Before establishing a connection, the backend verifies the endpoint’s address, port, and debug path to ensure it uses the local page endpoint specified by the caller.

When establishing a connection, the backend precisely matches the target based on the page identifier, rather than selecting candidate pages based on window titles. The session executes code within the page via \texttt{Runtime.\allowbreak evaluate}, waits for the Promise to complete, and returns the result by value. Script exceptions and execution results that cannot be converted to JSON values are both recorded as post-execution errors; if a disconnection, timeout, or unparseable response occurs after submission, the result is marked as unknown, and the request is not automatically retried.

\subsubsection{Electron Gadget Bridge}

For Electron hosts that meet the loading criteria but do not provide a page entry point via CDP, the system offers optional Gadget bridging. The Rust backend connects to authenticated Frida Gadget sessions via a Python helper, verifying the target device, process instance, and host configuration—rather than relying solely on a PID for access. The target-side helper script uses the host’s libuv scheduling mechanism to route requests into its native V8/Node.js execution environment; after obtaining the Electron main process interface, it then discovers \texttt{webContents} and calls its page execution interface. Therefore, script scheduling in the native bridge and the subsequent execution of the enhancement program within the page are two sequential steps.

The page identifier includes the target scope and verifies the current page by combining the \texttt{webContents} identity, URL, and main frame information. The Node.js-side acknowledgment carries the request ID and process information; the assistant reads from the log starting point of this request and matches it against the acknowledgment to avoid mistaking historical output for the current result. Native asynchronous handles and callbacks retain their references during use; script resources are released only after a close operation obtains the corresponding acknowledgment. This bridging relies on available host output symbols and a matching ABI; if these conditions are not met, it returns “not supported.”

\subsubsection{WKWebView Bridge}

The WKWebView backend reuses Frida session communication, but page discovery and evaluation utilize WebKit’s native interfaces. The macOS host adaptation traverses the window view tree on the main thread, selects the WKWebView within it, and organizes the target based on the view identity and URL; it re-locates and verifies the corresponding view with each execution. Page code runs in the main frame’s page content environment via \texttt{callAsyncJavaScript}, and return values and exceptions are encapsulated within the page before being passed back via native completion.

WKWebView’s execution relies on the main thread and asynchronous callbacks; therefore, a timeout on the control end does not necessarily mean the operation has been canceled. Tasks that have not yet begun are discarded; tasks that have already begun are marked as having an unknown result, and the necessary callback and resource information is retained to prevent false positives indicating that cleanup has been completed.

The CDP approach is intended for macOS and Windows hosts with corresponding page endpoints, while the Gadget and WKWebView bridging described in this section currently uses a macOS target configuration. Therefore, the fact that the control end can be built or use CDP on another platform does not imply that all host bridging has been ported. Specific entry points and version requirements for each application are reported separately in the evaluation section; experimental probes or interface declarations are not considered formal backend capabilities.

\subsection{Preparing a Derived Application}
\label{_sec:implementation-preparation_}

The previous section described the implementation of runtime integration; this section explains the implementation of pre-launch bootstrapping.
\texttt{prepare} The process reads the host’s package layout and main entry point, creates a copy of the derived application with a distinct identity, and installs Bootstrap within that copy, while leaving the original application unchanged.
The current implementation covers common Electron CommonJS packaging formats on macOS Apple Silicon, including directories, ASAR files, and nested ASAR entry points within directories. These packaging formats share the same preparation process, eliminating the need to write separate injection logic for each application; for entry points that cannot be recognized or are unsupported, the preparer will stop and report the reason.

The preparation process employs two strategies based on the entry point format. For CommonJS text entry points that can be parsed directly, a syntax parser is used to determine the positions of the file header and directive preamble, after which the bootstrap loading statement is inserted. The insertion position preserves the original file header and directives such as strict mode, avoiding the simple placement of the bootstrap statement at the beginning of the file that would alter the original entry point’s semantics. Modifications are limited to the entry point and necessary bootstrap resources within the derived copy; subsequent enhancements are not written to this entry point.

For entries nested within archives or handled by a host loader, the preparer employs a main module delegation strategy. The new entry file first loads Bootstrap, then initiates the original entry via Node.js’s main module loading mechanism. This strategy requires no modification to the original entry’s content and is therefore also suitable for entries where direct text insertion is not possible. This differs from the simple \texttt{require} call: the original entry point still runs as the main module, and its relative dependencies are resolved according to their original locations.

Upon completion of preparation, the preparer generates a preparation record that stores the derived instance identifier, entry point strategy, control-end address, and the status of key resources at the time of preparation completion for subsequent verification. The authentication materials and runtime logs required for Bootstrap communication are stored in the derived instance directory outside the repository. Subsequently, the macOS helper script updates the integrity records of relevant resources in the copy and signs the copy locally; finally, a pre-check confirms that Bootstrap can be loaded. The pre-check exits immediately after outputting a confirmation marker and does not continue executing the original business entry point; therefore, it is used solely to verify the preparation chain. Subsequent \texttt{dev} processes launch or connect to the corresponding derived instance based on the preparation record, and there is no need to re-execute the preparation process when updating the enhanced payload.

\subsection{Bootstrap Support for In-Application Iteration}
\label{_sec:implementation-bootstrap_}

Bootstrap launches alongside the derived application and, after completing instance authentication, provides the local control end with a channel for page discovery and enhancement execution. The application entry point only needs to be prepared once; subsequent modifications can be submitted to the selected page through this channel, allowing users to view the results within the original application without having to re-prepare the application package.

To apply modifications to the document currently visible to the user, the bootstrap code uses Electron’s \texttt{webContents} to discover the page and binds the target to the current connection session, page, and document generation to distinguish between different document loads within the same page. When the document is ready, the system writes a random marker into the page; this binding becomes invalid immediately after main page navigation, a change in the rendering process, or page destruction. Before executing each enhancement, the page first verifies this marker and then runs the payload, preventing erroneous execution caused by document changes occurring between the time the control end selects a page and the time the enhancement is actually executed.

Both requests and acknowledgments carry content version and target information; the control end accepts only results that match the current request. Connection restoration or the resending of an acknowledgment does not cause the original payload to be re-executed; when the result cannot be confirmed, the system retains an “unknown” status, and the caller decides the next action after verifying the page’s behavior. Version logging and diagnostic methods are described in Section \ref{_sec:implementation-diagnostics_}.

\subsection{Bundle Loading and Page Execution}
\label{_sec:implementation-payload_}

A bundle uses a versioned JSON manifest to declare the enhanced identifier, as well as the relative paths to JavaScript, HTML, and CSS resources. After the server reads and validates the manifest and resources, it assembles the payload into a program executable within the page. Input validation is used solely to ensure the format and paths are valid; it does not guarantee that the payload code is secure.

The program generator reuses the same Web runtime, serializing the enhanced identifier, HTML, and CSS into data, and assembling JavaScript into asynchronous function bodies. All backends share this runtime, eliminating the need to implement DOM assembly and update logic separately. The page runtime creates style nodes and an optional HTML root container, providing the payload with a reference to the root container and \texttt{onCleanup} a cleanup registration interface; if no HTML is present, the root container remains empty, and the payload can directly manipulate the original page.

The runtime registers nodes and cleanup callbacks based on the enhancement identifier. During updates, it first cleans up the old instance before executing the new payload and rejects overlapping operations on the same identifier. Cleanup affects only registered resources; even if a cleanup callback fails, the system continues processing the remaining resources and reports the error at the end. This mechanism supports replacing or removing individual enhancements without refreshing the page.

\subsection{Revision Management and Execution Feedback}
\label{_sec:implementation-diagnostics_}

The dev workflow uses the manifest and resource content digests as version identifiers to map a single file modification to a subsequent page execution. If a file changes or becomes temporarily invalid during reading, the system reports a source error and does not submit the payload; once a stable candidate is obtained, it is applied to the selected page in sequence. Within the same dev run, versions that have already been attempted will not be executed again due to repeated file events. This mechanism prevents mixed content from being submitted to the page during the saving process, but does not guarantee that updates to multiple files will be completed simultaneously.

Each submission records the request number, content version, and target, and saves the pending status before execution. Upon receiving a definitive result, the control end saves the corresponding acknowledgment; if the result is unknown, it retains the record and halts automatic updates. Even if the control end restarts, the user must verify the page status and explicitly confirm it before a new request can be initiated; the system will not automatically re-execute the original program. Thus, the caller can distinguish between a file that has been modified, a program that has been executed, and an execution result that is unknown.

Execution feedback is also used to guide subsequent modifications. In the Bootstrap path, the control end converts matching error stacks—based on the request, version, and the user script’s position within the generated program—into row and column numbers in the enhanced source file and associates them with the corresponding console logs; if no match is found, the original error is retained. Source errors, execution failures, and unknown results are reported separately, allowing the caller to correct the payload, verify the page, or end the current iteration. Stopping `dev` only terminates file monitoring and the communication connection; it does not close the application, undo the current page effects, or save this enhancement as a feature that automatically loads when the application starts.

\section{Reference Enhancements and Evaluation}
\label{_sec:evaluation_}

This section examines, through real-world host verification and reference-enhanced examples, whether Injectra can establish page execution channels in different applications, reuse enhanced execution mechanisms, and support in-application modification workflows. The evaluation combines system verification with example demonstrations, focusing on the capabilities actually provided by the system and their applicable conditions—a approach consistent with toolkit research, which selects evaluation evidence based on research questions \cite{ledo2018toolkit}.

\subsection{Evaluation Questions and Evidence Scope}
\label{_sec:evaluation-scope_}

The evaluation centers on three questions: Can the same set of page execution capabilities be reused across different hosts and access paths? Can these capabilities support presentation adjustments, interaction modifications, and enhanced instances generated by large language models? Do updates, cleanup, and execution failures have verifiable boundaries? These three questions correspond to access reuse, the scope of enhanced expression, and iterative control, respectively, and are not intended to estimate Injectra’s overall compatibility with all desktop software.

The evaluation data is derived from validation records retained during system development and deployment, primarily on macOS Apple Silicon, with supplementary validation conducted concurrently on Windows. Software versions, entry conditions, and validation objectives varied across different batches; some attempts also included method refinements and follow-up re-validation. Therefore, this section reports results based on original batches and specific combinations and should not be regarded as a randomized sampling experiment with a predetermined protocol. The basic unit of observation is the combination of application version, platform, access path, and selected page; repeated requests, other versions of the same application, and independent test hosts are not counted as new third-party applications.

Page execution verification uses custom payloads to check DOM mounting, style calculations, JavaScript interactions, updates under the same identifier, and resource cleanup. In the official CLI route, auxiliary tools are responsible for asserting page behavior, while payloads are still submitted by Injectra. Enhanced references are separately examined for the scope of their layout, control, or text modifications. The former verifies the execution channel, while the latter verifies the localized effects of specific enhancements; these two types of evidence are not interchangeable.

\subsection{Validation across Hosts and Access Paths}
\label{_sec:evaluation-hosts_}

Table \ref{tab:validation-overview} summarizes the evidence at each level. The first round of the macOS CDP survey identified 40 relevant candidates from native application packages; actual launch attempts involved 26 applications, totaling 31 attempts, with 20 applications ultimately completing all page validation checks. Of these, 18 used baseline payloads consisting of HTML, CSS, and JavaScript; the other two, when page policies restricted HTML strings, switched to generic payloads that constructed the DOM using JavaScript to complete verification. The remaining candidates—including those for which no entry point was obtained, those with unconfirmed isolation conditions, and those that failed to launch—are not grouped with post-execution failures and are not included in the calculation of the overall compatibility rate.

Subsequent records provided evidence for different versions and integration paths. WorkBuddy and Youku completed CDP verification after meeting the data directory requirements of their respective hosts; Douyin 8.7.0 completed execution and cleanup verification via an Electron Gadget page bridge; and Codex Switcher 0.2.11 completed main page verification and isolation checks for other views via the WKWebView path. Verification of standalone Electron, CEF, NW.js, and Swift/WKWebView hosts further covered common behaviors and exception boundaries across different rendering architectures, indicating that shared execution mechanisms can be reused across frameworks without being tied to a specific type of client implementation.

On the actual chat page of Doubao 2.33.5, applying the standard validation payload generated two managed nodes; clicking the button created by the test payload caused the counter to change from 0 to 1. After updating the payload while maintaining the same enhancement identifier, the number of nodes remained at 2 and the counter reset to 0; upon removal, the number of nodes returned to zero. The page state corresponds to the CLI acknowledgments, successfully verifying the loading, replacement, and cleanup flows on this page.

Pre-launch bootstrapping used the same preparer to overwrite directories, ASARs, and nested ASAR fixtures, and successfully validated incremental updates on the QQ 6.9.86 main page and the Claude 4.23118.1 local shell page. The QQ validation further adopted a main module delegation strategy, successfully completing updates and cleanup for two versions of JS, HTML, and CSS in succession, while the hashes of the original archived and prepared resources remained unchanged. This demonstrates that the strategy supports continuous iteration without modifying the original entry-point content. The Claude validation covered incremental updates to the local shell page; the remote chat view and optional console logs were not included in this validation, and the relevant results were recorded separately according to the actual validation targets.

The Windows installation list contains 38 items. In the first round, 29 items obtained evidence of successful page validation; subsequent re-verifications cumulatively reached 35 items, covering various integration methods such as the native Rust CLI/CDP, the Gecko test client, and experimental Bootstrap/V8. Subsequently, NetEase Cloud Music 3.1.38 utilized the existing Windows CLI and standard CEF debugging entry points to complete verification of the interaction, update, and cleanup of its proprietary page payloads. Youku, Baidu Netdisk, and WeCom were not included in this round of follow-up re-verification; therefore, there are currently no page verification results for them, and they are not counted as integration failures. Windows results are reported separately and are not combined with records from different versions and integration paths on macOS.
\begin{table}[htbp]
  \centering
  \caption{ A hierarchical overview of existing validations. Each batch is reported by original version, platform, and access path; subsequent re-tests do not update earlier records. Independent hosts, real applications, and experimental entry points are used to provide different types of evidence and are not combined when calculating the overall compatibility rate. }
  \label{tab:validation-overview}
  \small
  \setlength{\tabcolsep}{4pt}
  \renewcommand{\arraystretch}{1.2}
  \begin{tabularx}{\linewidth}{@{}>{\raggedright\arraybackslash}p{0.23\linewidth}>{\raggedright\arraybackslash}p{0.42\linewidth}>{\raggedright\arraybackslash}X@{}}
  \toprule
      Evidence Levels & Validation Results & Scope and Notes on Validation \\
  \midrule
      First Round of macOS CDP Survey
  & 40 candidates identified; 26 applications actually launched, totaling 31 attempts; 20 applications completed page validation, of which 18 used baseline payloads and 2 used payloads that constructed the DOM via JavaScript
  & The remaining candidates include those for which no entry point was obtained, those with unconfirmed isolation conditions, and those that were not launched; these are not counted as payload execution failures \\
  \addlinespace
      macOS Post-Launch Integration
  & WorkBuddy and Youku completed CDP verification; Douyin completed execution and cleanup verification via Electron Gadget; Codex Switcher completed main page and view isolation checks via WKWebView.
  & Results were recorded separately by application version, selected page, and integration path \\
  \addlinespace
      Independent Host Verification
  & Electron 36.4.0, CEF 156.0.26, NW.js 0.119.2, and Swift/WKWebView hosts passed the corresponding execution and lifecycle checks
  & Used to verify common behaviors and exception boundaries across frameworks; not counted toward the number of third-party applications \\
  \addlinespace
      macOS Pre-Boot Verification
  & Directories, ASAR, and nested ASAR fixtures passed the preparation verification; the QQ 6.9.86 main page and the Claude 4.23118.1 local shell page passed the update and cleanup verification
  & Covers currently supported Electron CommonJS entry points; Claude’s remote chat view and optional logs were not included in this verification \\
  \addlinespace
      Windows Manifest and Re-verification
  & 38 items were included in the installation manifest; 29 items passed page verification in the first round, and a cumulative total of 35 items passed in subsequent re-verifications; the remaining 3 items were not subject to further re-verification
  & Includes the official CLI, Gecko test client, and experimental entry points; different integration forms are counted separately and are not included in the 35 official backends \\
  \addlinespace
      Gecko Exploratory Validation
  & Firefox 160 passed page validation; Thunderbird 159 partially passed under the default HTML string payload, and subsequent DOMParser assembly experiments yielded positive results
  & This is considered exploratory validation and has not yet been incorporated into the official CLI backend; different payload assembly formats are recorded separately \\
  \bottomrule
  \end{tabularx}
  \end{table}

\subsection{Reference Enhancements and Iterative Use}
\label{_sec:evaluation-enhancements_}

All enhanced instances used in this section are implemented via independent plugin libraries or external control programs; page object selectors and business logic remain within the enhanced payloads and do not enter the Injectra core. Table \ref{tab:reference-enhancements} lists the validation results and applicable scopes for three types of instances: presentation adjustments, reorganization of existing interface interactions, and LLM-assisted modifications. These examples are used to verify the types of functionality supported by the shared execution foundation; they do not constitute user research samples, nor are they used to calculate a unified task success rate.

Presentation adjustments include switching between pink, purple, and the default appearance in QQ, as well as theme enhancements for the Doubao quick-access window. Verification of the QQ home page simultaneously checks actual controls, calculates background color values, and identifies input nodes; upon reverting to the default appearance, the corresponding styles are restored. All page adaptations required for theme switching are included in the payload; no modifications to the injection core are necessary. This example demonstrates that the same execution capability can support stateful, localized presentation adjustments, rather than merely inserting static markers.

The Doubao thesis reading plugin further reorganizes the existing interface: it separates the thesis area from the chat overlay—which can be expanded, dragged, and scaled—while reusing the host’s original PDF and chat components. Within the recorded 975-pixel-wide viewport, the paper area expands from 575 pixels to 975 pixels; when the chat overlay is expanded, moved, or scaled, the paper width and the current reading position remain unchanged. Upon exiting, the original layout is restored, and the input node identity remains unchanged. This example validates the layout and interaction effects on a specific page but does not measure reading efficiency, nor does it verify the compatibility of all chat functions by sending actual messages.

\begin{figure}[htbp]
\centering
\begin{minipage}[t]{0.30\linewidth}
\vspace{0pt}
\includegraphics[width=\linewidth,viewport=170 788 510 1038,clip]{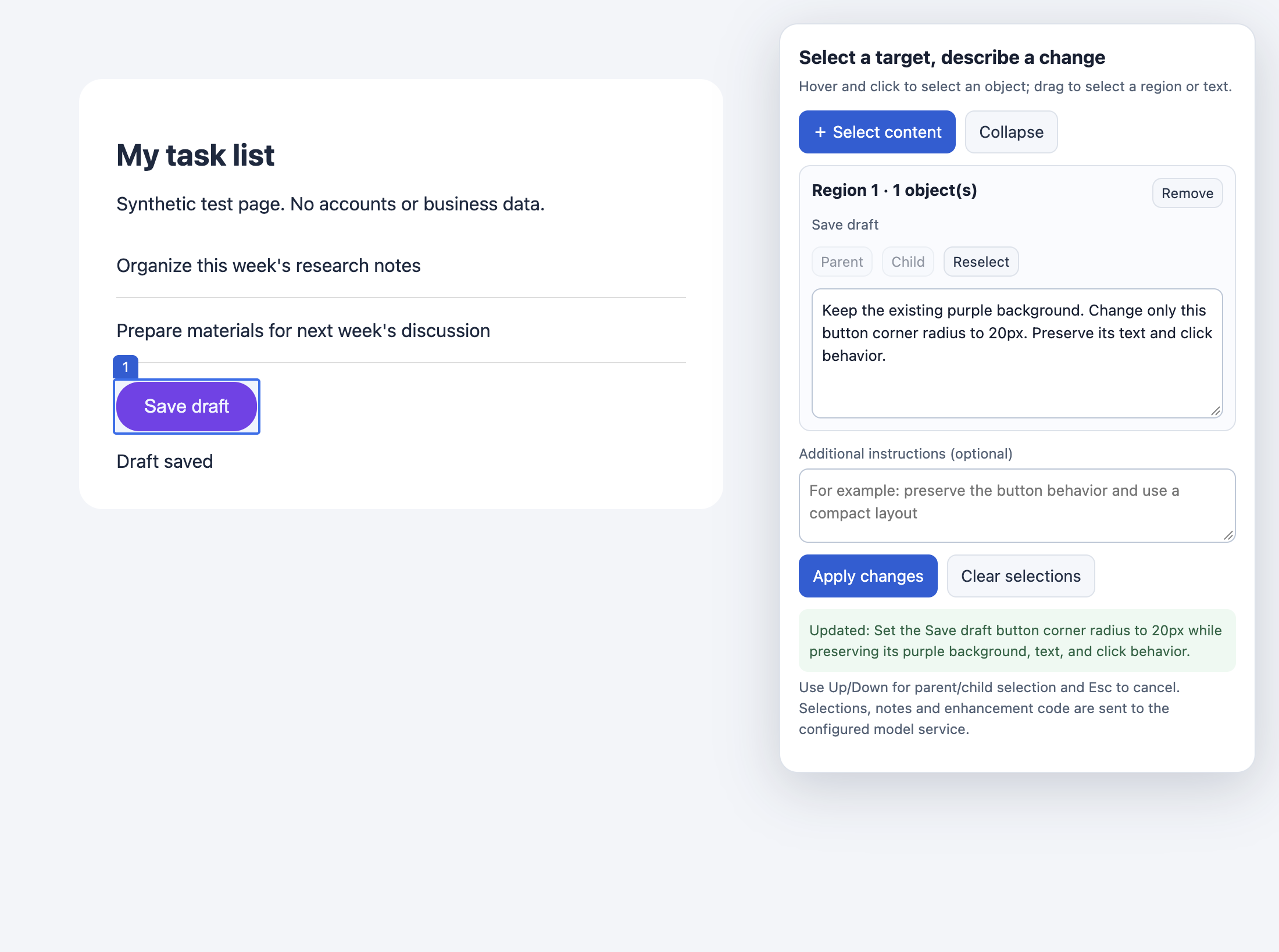}
\par\small \(a\) Target button and original save feedback
\end{minipage}\hfill
\begin{minipage}[t]{0.65\linewidth}
\vspace{0pt}
\includegraphics[width=\linewidth,viewport=1340 338 2160 1360,clip]{figures/llm-enhancement-example-en.png}
\par\small \(b\) Annotation and update feedback
\end{minipage}
\caption{ Examples of enhanced box annotations in an independent English Electron synthesis page. After two rounds of processing, the selected button retains its purple background and 20-pixel rounded corners, and the original save function remains functional upon verification. The two cropped sections are taken from the same original English verification image; the text in the image has not been redrawn or replaced. This image is used to illustrate the modification process. }
\label{fig:reference-llm-enhancement}
\end{figure}

The selection annotation tool places the code generation process on top of the shared execution layer. In a standalone Electron composite page, the test script submitted two rounds of requests via actual pointer and text input: the first round changed the save button to purple, and the second round changed the corner radius to 20 pixels while retaining the background. The recording shows that the external model service generated enhanced code and updated the page during both rounds of processing; the final computed styles met the requirements, the component remained a single instance, and the original Save button still generated a save confirmation. Figure \ref{fig:reference-llm-enhancement} shows a partial screenshot taken after the second round. The test page uses explicitly labeled synthetic data and does not contain user accounts or business content. This figure is from an independent English-language review; the synthetic page, annotation interface, and result feedback are all in English.

Another synthetic task involved modifying only the background of selected text, leaving text outside the selection area unchanged, and passing validation without an HTML container. Using the same data collection and execution methods, evidence corresponding to the target range and styles was also obtained through partial text modifications on real Doubao pages. These cases demonstrate that requirement collection, external code generation, and in-app execution can form a complete workflow; however, a small number of examples is insufficient to support an accuracy assessment for arbitrary natural language requirements. Model identifiers, prompts, and responses are all retained in the corresponding records; This section does not compare models, nor does it infer their internal mechanisms based on model names returned by the API.

\begin{table}[htbp]
  \centering
  \caption{ Refer to the enhanced reuse boundaries and validation results. Each instance reuses Injectra’s target page links, payload application, and updates with the same identifier; the first two items are cleaned up via the payload, and the generation process for LLM instances is organized by an external control endpoint. Synthetic pages are operated via scripts, and the results are not used to evaluate human task efficiency. }
  \label{tab:reference-enhancements}
  \small
  \setlength{\tabcolsep}{4pt}
  \renewcommand{\arraystretch}{1.2}
  \begin{tabularx}{\linewidth}{@{}>{\raggedright\arraybackslash}p{0.18\linewidth}>{\raggedright\arraybackslash}p{0.28\linewidth}>{\raggedright\arraybackslash}X@{}}
  \toprule
      Instances and Scenarios & Responsibilities Retained by the Enhancement Side & Verified Effects and Scope \\
  \midrule
      Theme Adjustments: QQ and Doubao Quick Windows
  & Page selector, theme color schemes, and switching controls; logic restored to original state
  & Color values calculated by specified controls meet expectations; input nodes are retained. QQ restores its original style after reverting to the previous state. Not all pages within the app were checked. \\
  \addlinespace
      Doubao Thesis Reading Layout
  & Positioning of thesis and chat components, overlay layout, and interactions; original layout restored
  & In a 975-pixel viewport, the thesis area expands from 575 to 975 pixels; the thesis geometry and reading position checked during overlay operations are maintained and restored upon exit. Reading efficiency and full chat functionality were not measured \\
  \addlinespace
      Two rounds of button modifications for LLM: standalone Electron composite page
  & Selected text annotations, DOM context, model invocation, and full version release; target selection and cleanup in generated code
  & After two rounds, the purple background with 20-pixel rounded corners is retained; the original save function works normally, and the data collection component remains a single instance. Used to validate the complete workflow; general generation accuracy is not estimated \\
  \addlinespace
      Local text modifications and continuous feedback
  & Text range capture, requirement transmission, and code highlighting; target text matching and cleanup
  & In the synthesized page, under empty HTML conditions, only the background of the selected text is modified; text outside the range is retained; the target range and styles on the Doubao page correspond. Dynamic objects or persistent cross-page positioning were not verified \\
  \bottomrule
  \end{tabularx}
  \end{table}

\subsection{Failure Modes and Execution Boundaries}
\label{_sec:evaluation-failures_}

The location where a failure occurs determines the conclusions that can be drawn from the results. The absence of a debugging page, the host lacking the required exports, or the copy failing to enter the target interface cannot be considered evidence that the payload has been executed. For example, the Gadget path in Electron45 lacks the required V8 export; although the local animation HTML candidate in Tencent Video was modified, the corresponding page markup and receipt did not appear during verification, so the presence of the application window alone cannot be used to conclude that the injection was successful. These cases document the entry points or loading boundaries for the corresponding versions and paths, but do not imply that other paths are infeasible.

The fact that a page can execute JavaScript does not imply that all forms of HTML assembly are applicable. In the first round of surveys, some page policies rejected HTML strings, while generic payloads that constructed the DOM using JavaScript still functioned. Thunderbird 159’s main interface only partially passed default HTML string validation; subsequent DOMParser assembly experiments yielded positive evidence, but it has not yet become the official CLI backend. This indicates that payload formats, page document types, and access methods must be reported separately; success at one level cannot be generalized to others.

Controlled host and protocol testing also covers incorrect target rejection, script exceptions, failed result encoding, unknown results due to timeouts, and prohibited automatic replay. The execution progress that can be confirmed varies by protocol: protocol errors without explicit page acknowledgments are conservatively marked as “unknown,” and results are not rewritten to achieve a neat classification. In synthetic scenarios using selection tools, annotations remain intact and editable even after a script exception occurs; when the result is “unknown,” the process requires stopping and verifying the current state. These validations focus on whether the post-failure state is interpretable; they do not imply that the side effects of any script can be cleaned up or undone.

\subsection{Summary of Findings}
\label{_sec:evaluation-summary_}

The above validations lead to three main conclusions. First, Injectra’s unified page execution contract and shared runtime can be reused across operating systems, host frameworks, and integration paths, validating the feasibility of a generic integration architecture. Second, the same execution foundation can deploy both theme and layout enhancements written by developers and support local modifications generated by LLMs, indicating that upper-layer enhancements do not need to reimplement host integration, and that page semantic adaptation can remain separate from the generic execution mechanism. Finally, identifier-based updates, managed resource cleanup, and execution feedback constitute a verifiable in-application iterative process, enabling enhancements not only to be injected but also to be continuously modified and verified within the original application.

These results demonstrate that Injectra integrates cross-application integration, enhancement execution, and iterative feedback into reusable infrastructure, providing a unified implementation pathway for LLM-generated code to enter real-world desktop software.
\FloatBarrier

\section{Discussion}
\label{_sec:discussion_}

\subsection{Retrofitting Malleability into Existing Software}

The enhanced example demonstrates that Injectra can perform software modifications directly within applications that users already rely on, without the need to rebuild replacement applications or migrate existing data, usage habits, or workflows. Theme adjustments, layout reorganization, and localized functional modifications act on interface presentation, interaction organization, and client-side logic, respectively, while continuing to reuse the original application’s document processing, input responses, and business components. Injectra thus provides a unified foundation for code integration and execution while preserving the flexibility of enhancements in terms of functional form and page adaptation.

This approach incrementally introduces software plasticity into existing Web UI clients: enhancements can be loaded, replaced, and cleared based on identifiers, allowing users to continuously view the results within the original application and initiate the next round of modifications. Compared to redesigning the software environment to be malleable, this approach does not require the application to migrate to a new data model or runtime environment; compared to external interface overrides, it directly utilizes the page’s internal DOM, styles, and event mechanisms, eliminating the need to maintain an overlay interface or input forwarding layer. Consequently, Injectra provides a transformation path for existing software that preserves the original runtime environment, supports multiple forms of enhancement, and enables sustainable iteration.

\subsection{Access Infrastructure and Enhancement Reuse}

The value of separating integration from enhancement lies in centralizing recurring host differences within a shared infrastructure. A single page integration can execute both handwritten payloads and support reading plugins or LLM modification tools, eliminating the need for the caller to implement protocol connections and page execution separately. Existing plugin libraries and independent control endpoints exemplify this approach to reuse: the latter uses existing library interfaces to handle requirement collection, code generation, and page updates, without needing to replicate the injection runtime.

This reuse does not imply that application semantics will automatically align. The same styles or scripts may correspond to different object selectors across different applications, and the content within a `Canvas` cannot be used to derive internal object semantics solely through the DOM interface. Therefore, it is necessary to evaluate separately whether the integration mechanism is reused, whether the enhancement payload requires adaptation, and whether the final functionality is correct. Current records support the feasibility of shared integration and execution, but have not yet quantified the integration costs, learning curve, or long-term maintenance workload for independent developers.

\subsection{Roles of Users and Language Models}

The selection and annotation tool demonstrates a clear division of labor between the LLM and Injectra: the model is responsible for converting user requirements into candidate enhancement code, while Injectra is responsible for applying the code to the original application page and providing execution feedback; the injection core does not need to handle requirement understanding or autonomous planning. Users specify the objects to be modified and their intent through selection and annotation, and the actual page results serve as the basis for subsequent verification and modification. Models, prompts, and interaction tools can be replaced independently, and both handwritten and generated payloads can share the same set of update, cleanup, and feedback mechanisms.

Clear information regarding requests, objectives, and results keeps this process controllable and verifiable. Successful execution indicates that the code has run on the target page, but whether it meets the requirements must still be determined by evaluating the page’s actual behavior; the selected area is used to express the intent of the modification and does not constitute a security boundary that restricts script behavior. By distinguishing between code generation, payload submission, and execution verification—and by preserving context information in the event of exceptions or unknown outcomes—Injectra provides a reliable foundation for human-in-the-loop iterative development within applications. As model capabilities and interaction methods evolve, this division of labor also offers a viable path for non-programmers to continuously refine everyday software using natural language.

\subsection{Deployment Trade-offs and Future Interfaces}

Runtime integration reduces the preparation steps for entry points but relies on the target already providing or allowing the establishment of corresponding channels; pre-launch guidance through derived copies supports subsequent iterations while introducing conditions such as package layout, signatures, and application launch. A unified interface cannot eliminate these differences. Preparing copies may also be affected by account data paths, shared containers, and software update methods; even if a specific page passes verification, this does not mean that a derived copy can unconditionally replace the original application as a daily usage environment.

These experiences suggest that if future hosts can provide clear extension entry points, page identity, and lifecycle covenants, they can reduce the uncertainty external tools face when adapting to existing software. The current implementation provides a set of reference abstractions and implementation methods, rather than an established cross-vendor standard. If interoperability is to be further standardized, independent clients or implementations must validate the common contract; if long-term personalized functionality is to be sustained, evidence of cross-application updates and long-term usability is required. Therefore, this study regards unified execution capabilities as a composable foundation layer, reserving generation methods, specific interactions, and long-term maintenance as separate research topics.

\section{Limitations and Conclusion}
\label{_sec:limitations-conclusion_}

\subsection{Limitations}

The current evaluation is primarily based on test samples obtained during the development and deployment process. The page results obtained so far do not yet cover all windows, business functions, or subsequent versions of the same application. Nevertheless, validation results across various hosts, platforms, and integration paths indicate that Injectra’s integration and execution mechanisms offer capabilities that can be reused across applications; however, additional adaptation may still be required for applications with protected entry points, unique page structures, or special runtime environments.

Whether the system can be integrated and what content it can modify depends on the entry points and page environment provided by the host. Current enhancements are primarily implemented through DOM, CSS, and JavaScript; therefore, they do not directly affect native controls, other frames, hidden views, or objects within Canvas. Linux and other hardware and host configurations have not yet been verified, and their usability cannot be inferred from the existing results. For hosts where an entry point cannot be established, resources are protected, or the required APIs are missing, corresponding integration mechanisms or adaptation implementations must be supplemented.

The payload runs as trusted code within the target page environment; the current system does not provide a security sandbox for arbitrary JavaScript. Cleaning nodes and registering callbacks can only handle registered resources and cannot undo persistence, network, or other external side effects; derived copies and isolated directories also cannot guarantee that account or system side effects have been completely isolated. The box-selection annotation tool, used as a verification reference example, relies on a user-configured external model service to process the UI context and generate code. The model may misinterpret user requirements, and the context may contain sensitive UI information; Injectra’s execution receipt can only confirm whether the code was executed, but cannot guarantee the correctness of the generated results nor eliminate the risk of information exposure.

The current workflow primarily supports loading for specific targets and iterations during the development phase. After the control end is stopped, the current page state can be retained; however, after a page refresh, window reconstruction, or application restart, the target must be reconfirmed, and the enhancement must be loaded according to the corresponding process. The system does not yet provide one-click consolidation and distribution of final enhancements, automatic uninstallation of Bootstrap, or automatic repair after application upgrades, nor does it guarantee indefinite operation or automatic recovery under any failure.

Existing evaluations have verified the actual effectiveness of the enhancement features by examining page states, user interactions, and control preservation in real-world applications, providing direct evidence of the system’s feasibility. These technical validations primarily address whether the features can be implemented and do not equate to improvements in user experience or work efficiency. Model services and page structures change over time; subsequent reproduction requires verification of the actual code, materials, services, and host versions. In the future, user studies and paired tasks can be further conducted on this basis to evaluate Injectra’s impact on development costs, user burden, and long-term usage benefits.

\subsection{Conclusion}

This paper proposes and implements Injectra, a general-purpose code injection and functionality enhancement system for existing Web UI desktop applications. The system decouples host integration from upper-layer enhancements, providing unified page execution capabilities through runtime integration and pre-launch guidance, and supports the enhancement, replacement, and inspection of applications via a shared runtime, explicit target binding, and execution feedback. Validation across multiple real-world hosts, along with LLM tools for theme customization, reading layout reorganization, and box-selection annotation, demonstrates the ability to reuse the same execution foundation across applications.

Against the backdrop of large language models capable of generating code based on natural language, Injectra further addresses the challenge of how generated code can be integrated into and executed within the software the user is currently using. It enables model-generated enhancements and developer-written payloads to share the same in-application execution foundation, while also providing a composable, unified entry point for various interactive tools and agents. Consequently, existing desktop software is no longer limited to relying solely on predefined features and settings set by developers but now has the foundation to evolve into a software environment that can be continuously extended for users. Future work can focus on broader host adaptation, enhanced long-term maintenance, and natural language interaction experiences to drive the implementation of this foundation in more real-world workflows.

\section*{Statement on the Use of Artificial Intelligence}
During the preparation of this manuscript, artificial intelligence tools were used to assist with structuring, initial drafting, and English translation. The author reviewed and edited all generated content and takes full responsibility for the final manuscript.

\begingroup
\raggedright
\bibliographystyle{plain}
\bibliography{references}
\endgroup

\end{document}